\documentstyle[sprocl]{article}

\def\be{\begin{equation}}
\def\ee{\end{equation}}
\def\bea{\begin{eqnarray}}
\def\eea{\end{eqnarray}}
\begin{document}

\title{NEUTRINO FLAVOR MIXING AND OSCILLATIONS IN FIELD THEORY}

\author{ELISABETTA SASSAROLI}

\address{Laboratory for Nuclear Science and Department of Physics \\ 
Massachusetts Institute of Technology, Cambridge, MA 02139, USA}
\address{Physics Department, Northeastern University, Boston, MA 02115}

\maketitle\abstracts{
The Lagrangian that is normally associated with Dirac neutrinos is 
analyzed in a complete and simple way through field theory. 
It is found that the elements of the neutrino
mass matrix are field strength renormalization constants and that the 
flavor fields can be applied directly to the one-particle energy states through
these rescaling factors. Moreover this Lagrangian describes neutrinos which
are in a state of mixed flavor at any space time point and therefore does not
describe the phenomenology of neutrino oscillations properly, except in the 
ultra-relativitic limit where such a description is possible.}

\section{Introduction}

Our discussion in the present paper will employ the following 
Lagrangian \cite{Elisa}%
$$
{\cal L}={\bar \psi }_e(i\gamma \cdot \partial -m_e)\psi _e+
{\bar \psi }_\mu
(i\gamma \cdot \partial -m_\mu )\psi _\mu -\delta 
({\bar \psi }_e\psi _\mu +{%
\bar \psi }_\mu \psi _e),\eqno(1.1) 
$$
which consist of two coupled Dirac neutrino fields $\psi _e$, $\psi _\mu $ 
with
masses $m_e$ and $m_\mu $ respectively. The interaction between the two
fields is provided through a lepton number violating term with a coupling
constant $\delta $. The model allows for exact diagonalization. Neutrino and
anti-neutrino flavor wave functions can be obtained  as matrix
elements of the quantized neutrino fields. For a general review about neutrino
physics, see for example \cite{Bil}$^-$ \cite{Kim}.

It is shown that the rotation matrix elements are field strength renormalization
factors. Therefore the fields that makes sense to quantize are the free renormalized
fields $\psi_1$ and $\psi_2$ obtained by the diagonalization of Eq. (1.1). 
Only for those fields it is 
possible to define creation and annihilation operators and one-particle 
states.
The fields $\psi_e$ and $\psi_{\mu}$ can be applied directly to the one-particle
energy states through the rotation matrix. 

Moreover, the conserved charge is the sum of the electron and muon charges which are not
conserved separately. Therefore the above Lagrangian describes neutrinos which are 
in a state of mixed flavor at any space-time point and they do not
describe the possibility to have only a given flavor at a given time, such
as for example at production. Different authors came to the same conclusions
through  different approaches from the one shown here 
\cite{Kan} $^-$ \cite{Bla}. 
 
The problem may be a deep one and associated with the possibility that neutrinos violate 
the equivalence principle \cite{Col,Gla}.  

\section{Field Theory Treatment}

By diagonalizing the Lagrangian defined in Eq. (1.1) we obtain the renormalized
masses
$$
m_{1,2}={\frac 12}[(m_e+m_\mu )\pm R],\eqno(2.1) 
$$
with 
$$
R=\sqrt{(m_\mu -m_e)^2+4\delta ^2}.\eqno(2.2) 
$$

Corresponding to the positive energy solutions $E_{1,2}=\sqrt{m_{1,2}^2+p^2}$, we
have the eigenfunctions
$$
\phi_{1,2}({\bf x},t)=\left( 
\begin{array}{c}
\frac 1{\sqrt{1+M_{1,2}^2}} \\ \frac{M_{1,2}}{\sqrt{1+M_{1,2}^2}} 
\end{array}
\right) {\frac 1{\sqrt{V}}}{\frac 1{\sqrt{2E_{1,2}}}}u_{1,2}(s,{\bf p})e^{i{\bf p}%
\cdot {\bf x}}e^{-iE_{1,2}t},\eqno(2.3) 
$$
where $s=1,2$ is the spin index, $u_{1,2}(s,{\bf p})$ are the Dirac spinors 
and
$$ 
M_{1,2}={\frac{m_\mu -m_e \pm R}{2\delta }}.\eqno(2.4) 
$$

The rotation matrix $U$ between the flavor fields $\psi _e$ $,\psi _\mu $
and the free fields $\psi _1$ and $\psi _2$ of renormalized masses $m_1$, $%
m_2$ respectively can be written in terms of the flavor vectors given by Eq. (2.3) 
as 
$$
U=\left( 
\matrix { 
{1 \over \sqrt {1+M_1^2}} & {M_1 \over \sqrt {1+M_1^2}}\cr
{M_1 \over \sqrt {1+M_1^2}} & -{1 \over \sqrt {1+M_1^2}}\cr
}\right) ,\eqno(2.5) 
$$
where we have used the fact that $M_1M_2=-1$.
It is possible to see that if we write the electron and neutrino fields $%
\psi _e$ $,\psi _\mu $ in terms of the fields $\psi _1$ and $\psi _2$
by means of  the rotation matrix $U$, 
the interacting Lagrangian given by Eq. (1.1) becomes uncoupled, i.e.  
$$
{\cal L}_D={\bar \psi }_1(i\gamma \cdot \partial -m_1)\psi _1+{\bar \psi }%
_2(i\gamma \cdot \partial -m_2)\psi _2.\eqno(2.6) 
$$
The fields $\psi_1$ and $\psi_2$ are the dynamical variables which have to be
quantized through the canonical anti-commutation relations. Only for them 
is it 
possible to define creation and annihilation operators and 
one-particle states.
The fields $\psi_e$ and $\psi_{\mu}$ can be applied to the energy one-particle 
states using the rotation matrix defined in Eq. (2.5). 
For example, the 
neutrino wavefunction associated with the one-particle state 
of defined energy $E_1$ ($|1_{{\bf p}s}>$) is 
$$
\psi _\nu ({\bf x},t)=\left( 
\matrix {\psi_e({\bf x}, t)\cr
\psi_{\mu}({\bf x}, t)
}\right) =\left( 
\matrix {<0|\hat {\psi}_e({\bf x}, t)|1_{{\bf p}s}> \cr
<0|\hat {\psi}_{\mu}({\bf x}, t)|1_{{\bf p}s}> \cr }\right) = 
$$
$$
=\left( 
\matrix { {1\over \sqrt {1+M^2_1}} \cr
{M_1\over \sqrt {1+M^2_1}}\cr
}\right) {\frac 1{\sqrt{V}}}{\frac 1{\sqrt{2E_1}}}u_1(s,{\bf p})e^{i{\bf p}%
\cdot {\bf x}}e^{-iE_1t}.\eqno(2.7) 
$$ 

This, being a plane wave, gives a stationary probability of finding a
neutrino at a given space-time point. However in any location inside the
volume V there is a probability equal to $({\frac 1{1+M_1^2}})$ of finding
the neutrino in the electron flavor and probability equal to $({\frac {M_1^2%
}{1+M_1^2}})$ of finding it in the muon flavor. 

The constants $({%
\frac 1{1+M_1^2}})$ and $({\frac{M_1^2}{1+M_1^2}})$ are field-strength
renormalization factors and give the probability amplitude for the field
operator $ \psi_e$ ($ \psi_{\mu}$) to create a one-particle
eigenfunction of definite energy.

To be able to describe neutrinos which are in a  superposition of
different energies, we need a superposition of one-particle states. 
A general state of positive
charge, momentum ${\bf p}$ and spin $s$ is given by 
$$
|\phi _{+}>=[Ab_1^{\dagger }(s,{\bf p})+Bb_2^{\dagger }(s,{\bf p})]|0>,%
\eqno(2.8) 
$$
where $A$ and $B$ specify the amount of each normal mode state of positive
energy present in the state $|\phi _{+}>$ and 
$$
|A|^2+|B|^2=1.\eqno(2.9) 
$$ 
However it is possible to show  that even the flavor wavefunctions associated with the state 
given by Eq. (2.8) describe neutrinos which are in a state of
mixed flavor at any space-time point \cite{Elisa,Elisa1}. 
Therefore they do not describe properly
the neutrino oscillation phenomenology, where it is assumed that only
one flavor is present at production. The standard neutrino oscillation
probabilities can be recovered by making some {\it ad hoc} ultra-relativistic
approximations in the wavefunctions.    

The generalization of the model described here to two component Majorana neutrinos and 
to the CKM matrix for the quark sector can be found respectively in Ref.[13]
and Ref. [14].

\section*{Acknowledgments}

The author would like to thank Prof.  Alan H. Guth for his constructive 
criticism. She would also like to acknowledge
fruitful discussions with the MIT atomic and molecular
interferometry group.

This work is supported in part by funds 
provided by the U.S. Department of Energy (D.~O.~E.) under 
cooperative research agreement $\#$DF-FC02-94ER40818.

\end{document}